\documentclass{llncs}
\usepackage{xy}
\xyoption{all}

\usepackage{amssymb}
\usepackage{amstext}
\usepackage{amsmath}
\usepackage{txfonts}

\newcommand{\indiv}{private}
\newcommand{\Indiv}{Private}

\newcommand{\truss}[2]{\underset{#2}{\stackrel{#1}{\rightarrow}}}

\newcommand{\Obj}{{\sf J}}
\newcommand{\Atr}{{\sf U}}
\newcommand{\Rat}{{\sf R}}
\newcommand{\Bin}{{\sf B}}
\newcommand{\bind}{b}
\newcommand{\bindd}{\beta}
\newcommand{\Del}{{\sf D}}
\newcommand{\del}{d}
\newcommand{\dell}{\delta}

\newcommand{\Prob}{{\rm Prob}}
\newcommand{\Sim}{{\rm s}}
\newcommand{\pthc}{\#}

\renewcommand{\to}{\xymatrix@C-.5pc{\ar[r]&}}
\newcommand{\ot}{\xymatrix@C-.5pc{& \ar[l]}}
\newcommand{\tto}[1]{\xymatrix@C-.5pc{\ar[r]^-{#1}&}}
\newcommand{\oot}[1]{\xymatrix@C-.5pc{&\ar[l]_-{#1}}}
\newcommand{\mono}{\xymatrix@C-.5pc{\ar@{>->}[r]&}} 
\newcommand{\epi}{\xymatrix@C-.5pc{\ar@{->>}[r]&}}
\newcommand{\mmono}[1]{\xymatrix@C-.5pc{\ar@{>->}[r]^-{#1}&}} 
\newcommand{\eepi}[1]{\xymatrix@C-.5pc{\ar@{->>}[r]^{#1}&}}
\renewcommand{\mapsto}{\xymatrix@C-.5pc{\ar@{|->}[r]&}}
\renewcommand{\mmapsto}[1]{\xymatrix@C-.5pc{\ar@{|->}[r]^{#1}&}}
\newcommand{\inclusion}{\xymatrix@C-.5pc{\ar@{^{(}->}[r] &}}
\newcommand{\iinclusion}[1]{\xymatrix@C-.5pc{\ar@{^{(}->}[r]^{#1}&}}
\newcommand{\dtto}[2]{\xymatrix@C-.5pc{\ar@<.875mm>[r]^{#1} \ar@<-.875mm>[r]_{#2}&}}

\newcommand{\DDD}{{\cal D}}

\renewcommand{\Bbb}{\mathbb}
\newcommand{\AAa}{{\Bbb A}}

\newcommand{\EEe}{{\Bbb E}}

\newcommand{\NNn}{{\Bbb N}}

\newcommand{\RRr}{{\Bbb R}}

\newcommand{\TTt}{{\Bbb T}}

\mathcode`\<="4268 
\mathcode`\>="5269 
\mathchardef\gt="313E 
\mathchardef\lt="313C 

 \def\pushright#1{{
    \parfillskip=0pt            
    \widowpenalty=10000         
    \displaywidowpenalty=10000  
    \finalhyphendemerits=0      
    \leavevmode                 
    \unskip                     
    \nobreak                    
    \hfil                       
    \penalty50                  
    \hskip.2em                  
    \null                       
    \hfill                      
    {#1}                        
    \par}}                      

 \def\qed{\pushright{$\square$}\penalty-700 \smallskip}

\newenvironment{prf}[1]{\begin{trivlist} \item[{\bf ~Proof}#1.]}%
{\qed\end{trivlist}}

\newcommand{\be}[1]{\begin{#1}}

\newcommand{\ee}[1]{\end{#1}}
\newcommand{\beq}{\begin{equation}}
\newcommand{\eeq}{\end{equation}}
\newcommand{\ba}[1]{\begin{array}{#1}}
\newcommand{\ea}{\end{array}}
\newcommand{\bea}{\begin{eqnarray}}
\newcommand{\eea}{\end{eqnarray}}
\newcommand{\bear}{\begin{eqnarray*}}
\newcommand{\eear}{\end{eqnarray*}}
\newcommand{\bpr}{\begin{prf}{}}
\newcommand{\epr}{\end{prf}}
\newcommand{\bprf}[1]{\begin{prf}{#1}}
\newcommand{\eprf}{\end{prf}}

\newtheorem{thm}{Theorem}
\spnewtheorem*{onlythm}{Theorem}{\bf}{\itshape}

\newtheorem{cond}{}[thm]

\newtheorem{prenumb}[thm]{\hspace{-1ex}}

\title{Dynamics, robustness and fragility of trust}

\author{\author{Dusko Pavlovic\thanks{Supported by ONR and EPSRC.}\\%
\institute{Kestrel Institute and
Oxford University}
\email{\small Email:~dusko\char64\{kestrel.edu,comlab.ox.ac.uk\}%
}}
}

\begin{document}

\maketitle

\begin{abstract}
Trust is often conveyed through delegation, or through recommendation. This makes the trust authorities, who process and publish trust recommendations, into an attractive target for attacks and spoofing. In some recent empiric studies, this was shown to lead to a remarkable phenomenon of {\em adverse selection}: a greater percentage of unreliable or malicious web merchants were found among those with certain types of trust certificates, then among those without. While such findings can be attributed to a lack of diligence in trust authorities, or even to conflicts of interest, our analysis of trust dynamics suggests that public trust networks would probably remain vulnerable even if trust authorities were perfectly diligent. The reason is that the process of trust building, if trust is not breached too often, naturally leads to power-law distributions: the rich get richer, the trusted attract more trust. The evolutionary processes with such distributions, ubiquitous in nature, are known to be robust with respect to random failures, but vulnerable to adaptive attacks. 
We recommend some ways to decrease the vulnerability of trust building, and suggest some ideas for exploration. 
\end{abstract}

\section{Introduction}
\subsubsection*{Background.}
In analyzing security protocols, we often reason under the assumption that a protocol participant, say Alice, is honest. This assumption simply means that Alice acts just as prescribed by the protocol, and does not engage in any other available runs. Such an assumption is sometimes justified, and sometimes not. When this assumption about Alice is made by another protocol participant, say Bob, then we say that Bob {\em trusts\/} Alice. The notion of protocol, according to which Alice is trusted to behave, is understood in the broadest sense of the word, as a general constraint on participants' behavior. E.g., a conversation protocol may consist of the requirement that the participants speak the truth, and Bob may trust Alice in that sense. While Alice's statements may be true or false, Bob's trust may go through many shades of gray, and through some nuances of other colors. Trust is dynamic, and can be many-valued. But note that it does not depend on any rules outside the specified protocol: e.g., a bank robbery protocol may involve a requirement that the robbers do not shoot at each other, so Bob may trust Alice in that sense. In any case, we write $B\truss{\Phi}{r}A$, where $B$ob is the trustor, $A$lice is the trustee, $\Phi$ is the entrusted protocol (constraint, property), and $r$ is a trust rating, which quantifies the level of trust.

In practice, this general notion of trust is usually restricted to some special cases:
\begin{itemize}
\item in web commerce, the seller and the buyer are trusted to act according to the established exchange protocols; more generally, trust plays an essential role in web services and service-oriented architectures at large;
\item in access control, various types of principals (people, machines, services, channels) may entrust each other with various actions, or they may delegate authorities for such actions to each other \cite{Benantar:AC,LampsonB:AC};
\item in public key cryptography, it is useful to view keys as principals\footnote{Statically, two principals knowing the same keys are indistinguishable by cryptographic means. Dynamically, they may be distinguishable, e.g., by the fact that at some previous moment only one of them knew a particular key. Nevertheless, it is often useful and convenient to treat the keys as first-class citizens of cryptographic protocols, and to distinguish the principals only when necessary.}, and to view the key hierarchies as trust relationships \cite{BBK,LevienR,MaurerU:Trust,ReiterM:Metric},
\item various peer-to-peer and business-to-business transactions are based on trust, and the corresponding networks require various types of trust infrastructure \cite{Guha-Tomkins,Garcia-Molina:eigentrust,Karabulut,Garcia-Molina:Taxonomy}.
\end{itemize}

When social relations need to be analyzed, the modeling techniques often proceed from two different points of view: local and global. E.g. in economics, when the questions of risk and utility are analyzed from a local point of view, they subsume under microeconomics; when they are analyzed from a global point of view, they fall under {\em macro\/}economics. Analyses of trust fall into two roughly analogous categories. 

{\em Local\/} analyses of the trust relationship  $B\truss{\Phi}{v}A$ are largely concerned with the logics of $\Phi$, i.e. with the reasoning whereby the trustor $B$ conveys or justifies entrusting the trustee $A$ with $\Phi$. As explained above, the trust statements internalize principals' beliefs and interactions, and vary through different forms of uncertainty, which lead to nonstandard logical features and formalisms. The examples of this kind of approach include \cite{SassoneV:Trust,GuttmanJ:Trust,JosangA:Subjlog,LampsonB:AC,Ninghui:Trust,Ninghui:JACM}. E.g., when trust is analyzed in strand spaces \cite{GuttmanJ:Trust}, a trust relationship $B\truss{\Phi}{v}A$ is viewed on the level of a single send-receive interaction, where $A$ is the sender and $B$ the receiver. This interaction is annotated by a statement $\Phi$, which the receiver $B$ requires, and the sender $A$ guarantees. By sending the message, $A$ asserts $\Phi$; when he receives the message, $B$ assumes $\Phi$. The statement that $B$ trusts $A$ thus means that $B$ relies on $A$ for $\Phi$.

On the other hand, the {\em global\/} analyses of trust usually look at the {\em trust networks\/} spanned by the trust relationships $B\truss{\Phi}{v}A$ between the members $A,B\ldots$ of some set of principals. While the local analyses focus on the logics of the entrusted properties $\Phi$, the global analyses focus on the network structure and traffic dynamics leading to trust, and arising from it. The examples include \cite{BlazeM:Decentralized,Guha-Tomkins,LevienR,MaurerU:Trust,ReiterM:Metric}. In some cases \cite{Guha-Tomkins}, the entrusted properties are left implicit, because all trust relationships of interest concern the same $\Phi$ (e.g., $\Phi(A) =$ "$A$ is a reliable merchant" or  "$A$'s keys are not compromised"). In other cases, the analyzed trust concerns boil down to two \cite{BBK,LevienR}, or four \cite{MaurerU:Trust} types of trust relationship, which are simply annotated by different types of arrows. Although the logics of trust have also been investigated in the context of  trust networks  \cite{JosangA:PGP,JosangA:Trustnet06}, many  basic questions about trust dynamics remain widely open even when there is only one entrusted property.

\subsubsection*{Summary of the paper.}
We analyze dynamics of trust networks. It is driven by the users, who are trying to decide which web merchants to buy from, or in the Public Key Infrastructure model, which keys to use. The security problem for the user is that a trust authority, which she consults for trust recommendations, may be corrupt, just like any merchant, or any key. In order to decide which merchants to trust, the user must decide which recommenders to trust. And in order to decide which recommenders to trust, she must try some of the recommended merchants. The problem of the chicken and the egg arises. In order to protect herself, the user must not accept the trust recommendations passively, but  needs to build up her {\indiv}  trust vectors, perhaps using some public recommendations on the way. While the public recommendations cover a broader range of trust objects and interactions, {\indiv} trust vectors are less likely to be corrupt.

In section \ref{Modeling}, we present an abstract model of public trust networks. In section \ref{individual}, we analyze dynamics of the {\indiv} trust building and updating. In section \ref{Conclusions} we spell out the conclusions. In section \ref{Combining}, we discuss the applications, and propose some ideas how to combine {\indiv} trust vectors with public recommendations, towards more reliable trust decisions.

Trust networks, as presented in section \ref{Modeling}, consist of two components, echoing the distinction between the direct and indirect trust. This distinction is a common feature of most of the trust network models encountered in the literature \cite{BBK,LevienR,MaurerU:Trust,ReiterM:Metric}. Enriched with additional features, our model can be instantiated to these richer models. However, in order to present a picture simple enough for our analyses, we also show how to absorb, in a matrix form of a trust network, the chains of indirect trust, which is conveyed from one recommender to another, together with the direct trust, which is conveyed from the recommenders to the shops. 

In section  \ref{individual}, we show that, under reasonable assumptions, the process of trust building asymptotically converges to a power-law distribution of trust vectors. This means that trust distributions have heavy tails of highly rated {\em trust hubs}. One consequence is that trust distributions are thus resilient to random perturbations. Another consequence is that they are vulnerable to adaptive attacks on their trust hubs. The proviso is that the cheaters do not wait too long with their deceit. In our trust model, this proviso is represented by the assumption that, the more trust a principal accumulates by acting honestly, the less likely it becomes that he will turn out to be dishonest.

The conclusions are spelled out in section \ref{Conclusions}. Our analysis of trust dynamics applies both to  users' private trust vectors, and to recommenders' public recommendations. Since the latter are open to attacks, and turn out to obey the vulnerable power law distributions, they should not be directly used for trust decisions, but combined with the {\indiv} trust values. This suggestion is supported by the empiric evidence that the public trust vectors are often actually subverted\cite{EdelmanB:adverse}.  In section \ref{Combining}, we sketch some methods to combine public and private trust vectors, that need to be explored and evaluated in future research.

\section{Modeling trust networks}\label{Modeling}
In many communication networks, it is impossible, or unfeasible to fully authenticate and authorize all interactions. {\em Trust networks}\/ provide a supplementary service of partial authentication or authorization. In many cases, authentication is bootstrapped by incrementally strengthening trust. 

We begin by an informal description of the conceptual components of a trust network, and later provide the formal definitions. To determine thoughts, we first present the special case of a web shopping scenario. A shopper visits a virtual network of web merchants. If she has no prior experience with it, she can seek advice from some recommenders. Denote the set of merchants by $\Obj$ and the set of recommenders by $\Atr$. The recommenders record and process the merchant ratings, submitted by the users after their interactions with the merchants. From these ratings, the recommenders derive their  recommendations, and publish them as trust certificates. A trust certificate $c$ is represented by an expression in the form $u\truss{c}{r} i$, where $u\in \Atr$ is a recommender, $i\in \Obj$ a merchant, and $r$ is the trust rating in a previously agreed rating scale $\Rat$. A {\em recommendation network\/} $\AAa$ is spanned by such certificates.

In addition to the merchant recommendation certificates $u\truss{c}{r} i$, a recommender $u$ may issue the {\em endorsement certificates} $u\truss{e}{r} v$, where $v$ is another recommender. The endorsement certificates span an {\em endorsement network} $\EEe$. The endorsement chains, represented by the paths through the endorsement network, allow analyzing the subtle problems of transitivity of trust.

We call {\em trust network\/} a pair $\TTt = <\AAa,\EEe>$, where $\AAa$ is a recommendation network, and $\EEe$ is an endorsement network over the same set $\Atr$ of recommenders. Trust networks can be presented in many slightly different ways, but they all model the public infrastructure of trust.

Besides the shopping scenarios, trust networks also model the Public Key Infrastructures (PKI). In this interpretation, the trust authorities $u\in \Atr$ are not recommenders, but simply keys. The endorsements $u\truss{e}{r} v$ between them are now the {\em delegation certificates}. The objects of trust $i\in \Obj$ do not represent the web merchants any more, but the bindings between some principals' identities and their keys. A recommendation $u\truss{c}{r}i$ is now a {\em binding certificate} for $i$, signed by $u$. More details about this interpretation, and about other presentations of trust networks, can be found in \cite{BBK,LevienR,MaurerU:Trust,ReiterM:Metric}.

We proceed with the formal definitions.

\subsection{Recommendation networks}\label{Recommendation}
A {\em recommendation (certificate)\/} network  is an edge-labelled bipartite graph  
\bear
\AAa   & = &  \big(\Rat \oot{\bind} \Bin \tto{<\partial,\varrho>} \Atr\times\Obj \big)
\eear
where 
\begin{itemize}
\item $\Obj$ is a set of {\em objects},
\item $\Atr$ is a set of  {\em trust authorities}, or {\em recommenders},
\item $\Bin$ is a set of {\em certificates}, or {\em recommendations}, and
\item $\Rat$ is a set of {\em values}, usually an ordered rig, where the {\em trust ratings} are evaluated.
\end{itemize}
A recommendation (certificate) $u\truss{c}{r}i$ is thus represented by an edge $c\in\Bin$ of the graph, with the source node $\partial(c) = u$ and the target node $\varrho(c)=i$. The value $r = \bind(c)$ is the trust rating assigned to $i$ by $u$'s recommendation $c$. The same recommender $u$ may issue several recommendations $c_1, c_2\ldots$  for the same object $i$, with the same or different trust ratings; he may also revoke some of them. The use of these multiple recommendations may be regulated by various policies, summing up or averaging the ratings, validating only the last one, and so on. For simplicity, in the present paper we assume that each trust authority takes care for this, and publishes at each point in time at most one recommendation for each object, which sums up (or averages) all its valid recommendations for that object. This allows us to conveniently reduce recommendation networks to matrices $A = (A_{ui})_{\Atr\times \Obj}$, where 
\bear
A_{ui}  & = &  \sum_{u\truss{c}{} i} \bind(c)
\eear
The summation is taken in the {\em rig\/} structure of $\Rat$. A {\em rig\/} $\Rat = (\Rat,+,\cdot,0,1)$ is a "ring without the negatives". This means that $(\Rat,+,0)$ and $(\Rat,\cdot,1)$ are commutative monoids\footnote{Rigs are sometimes called {\em semirings}. But it seems more reasonable to call semiring an algebra $\Rat = (\Rat,+,\cdot)$ where $(\Rat,+)$ and $(\Rat,\cdot)$ are semigroups, satisfying $a(b+c) = ab+ac$.} satisfying $a(b+c) = ab+ac$ and $0a=0$. The typical examples include natural numbers $\NNn$, non-negative reals $\RRr_+$, but also distributive lattices, which in general cannot be embedded in a ring.  For concreteness, we shall work mostly with $\Rat = \NNn$ or $\Rat = \RRr_+$, i.e. assume that the trust ratings are nonnegative real numbers. 
It should be noted, however, that  in some concrete applications more general rigs are needed, e.g. of polynomials or affine functions over $\RRr_+$.

On the other hand, if the idea that our trust ratings have no upper bound seems strange, the reader can translate all our constructions to the interval $\Rat = [0,1]$, with the rating function $\bindd : \Bin \to [0,1]$ set to 
\bear
\bindd(c) & = &  1-2^{-\bind(c)}
\eear
The inverse transform is $\bind(c) = -\log_2\left(1-\bindd(c)\right)$. Being able to switch between these two equivalent views is useful because each simplifies different aspects of rating: the ratings over $\RRr_+$ are simpler when there are several parallel recommendations, which we want to add up, whereas the ratings over $[0,1]$ are simpler when there is a chain of recommendations, and we want to multiply them. 

\paragraph{Remarks.} While $\RRr_+$ and $[0,1]$ are just special cases of $\Rat$, one could also raise the opposite objection, that they are needlessly general, since most real systems accept and generate their ratings over some very simple lattice (such as $\star \lt \star\star \lt \star\!\star\!\star$). But data analysis is never performed within that lattice. E.g., if the ratings are derived from users' feedback, then they usually need to be balanced, before they are entered in the same data set, because some users tend to rate more generously than others. In some other cases, the ratings need to be normalized into a given interval. So the rig operations are usually necessary. On the other hand, in relational data analysis, $\Rat$ is the boolean algebra $\{0,1\}$, and the full ring structure is not given. So rigs are a reasonable compromise for general explorations. 

\subsection{Endorsement networks}
We model an {\em endorsement\/} network as an edge-labelled graph 
\bear
\EEe   & = &  \big(\Rat \oot{\del} \Del \tto{<\partial,\varrho>} \Atr\times\Atr \big)
\eear
where an endorsement (certificate) $u\truss{e}{r}v$ is represented as element $e\in\Del$ with $\partial(e) = u$ and $\varrho(e)=v$. The trust rating $r = \del(c)$ this time quantifies $u$'s endorsement of $v$. Like before, we reduce this network to a matrix $E = (E_{uv})_{\Atr\times \Atr}$, where
\bear
E_{uv} & = & \sum_{u\truss{e}{} v} \del(e)
\eear

Abstractly, an endorsement network is similar to some of the popular network models, used for analyzing  protein interactions, the Web, social groups, etc. (Cf. \cite{Langville06google-book,Newman:networks-book}, and the references therein.) Its dynamics can always be analyzed in terms of promotion, discussed in \cite{PavlovicD:CSR08}. In that paper, path completions were introduced to allow analyzing the multi-hop network interactions within a simple matrix framework. Here, they will allow us to analyze chains of trust in a similar framework. 

\subsection{Path completions of endorsement networks}
To some extent, trust is transitive: if $u$ trusts $w$, and $w$ trusts $v$, then $u$ can accept some reliance on $v$. But not too much. Depending on the level of risk, and the presence of alternatives, $u$ might prefer to avoid indirect trust. And in any case, it would be unwise for her to rely upon someone removed from her by 20 trustees of trustees of trustees\ldots Can we capture such subtleties without complicating the model?

A {\em chain\/} or {\em path\/} $u\stackrel{e}{\rightarrow} v$ in an endorsement network $E$ is a sequence of  links 
$u \stackrel{e_1}{\rightarrow} w_1 \stackrel{e_2}{\rightarrow}w_2
\rightarrow \cdots \stackrel{e_n}{\rightarrow} v$.
Given an endorsement network $E$, we would like to define another such network $E^\pthc$ over the same set of recommenders, but with the chains of the endorsement certificates as the new endorsement certificates. The naive idea is to simply take all finite chains of network links as the new network links; i.e., the paths through the old network become the links of the new network. The new network is then closed under composition: each path from $u$ to $v$, as a composite of some links through other nodes, corresponds to a link from $u$ to $v$. This amounts to generating the free category over the network graph. 

Unfortunately, besides the trust dissipation, described above, this kind of closure destroys a lot essential information in all networks, just like the transitive closure of a relation does. E.g., in a social network, a friend of a friend is often not even an acquaintance. Taking the transitive closure of the friendship relation 
obliterates that fact. Moreover, the popular "small world" phenomenon 
suggests that almost {\em every two people\/} can be related through 
no more than six friends of friends of friends\ldots So already adding 
all paths of length six to a social network, with a symmetric friendship 
relation, is likely to generate a complete graph. In fact, the average 
probability that two of node's neighbors in an undirected graph are 
also linked with each other is an important factor, called {\em 
clustering coefficient\/} \cite{Watts-Strogatz}. On the other hand, in 
some networks, e.g. of protein interactions, a link $u\rightarrow v$ 
which shortcuts the links $u\rightarrow w\rightarrow v$ often denotes a 
direct {\em feed-forward} connection, rather than a composition of the 
two links, and leads to essentially different dynamics. For all these reasons, only some "short" paths can be added to a network. This is assured by penalizing the compositions.

As mentioned above, the ratings within $\Rat = [0,1]$ are more convenient for analyzing the chains of trust, so we use it in the next couple of definitions.

\be{definition}
For a given endorsement network $\EEe  =  \big([0,1] \oot{\dell} \Del \dtto{\partial}{\varrho} 
\Atr\big)$,  a trust threshold $\eta \in [0,1]$, and a composition penalty $\epsilon \in [0,1]$, we define the {\em  path completion} to be the network
\bear
\EEe^{\pthc} & = &  \big([0,1] \oot{\dell} \Del^\pthc \dtto{\partial}{\varrho} \Atr\big)\mbox{ where}\\
\Del^{\pthc} & = & \{e \in \Del^+\ |\ \dell(e)\geq \eta \}\mbox{ and}\\ 
\dell\big(u_0 \stackrel{e_1}{\rightarrow} u_1 \stackrel{e_2}
{\rightarrow}u_2\rightarrow \cdots  \stackrel{e_n}{\rightarrow} u_n\big) & 
= & \epsilon^{n-1} \prod_{k=1}^n \dell(e_k)
\eear
with $\Del^+$ denoting the set of all nonempty paths in $\EEe$, i.e. $n\geq 1$.
\ee{definition}

\paragraph{Remark.}
A path-complete network $\EEe^\pthc$ is closed under the compositions of high-trust endorsements, but not under the compositions which fall below the trust threshold. It is not hard to see that the path completion is an idempotent operation, i.e. $\EEe^{\pthc\pthc} = \EEe^{\pthc}$, but that it may fail to be a proper closure operation, because the endorsements $e \in \EEe$ such that $\dell(e)\lt \eta$ are not in $\EEe^\pthc$, so that generally $\EEe\not \subseteq \EEe^\pthc$.

\subsection{Completions of trust networks}
At the final step of completing a trust network, we bring the information captured in it into a more manageable form by folding the completion of the endorsement part into a new recommendation network. The trust matrix, extracted from this recommendation network in the same way as before, now captures not only the direct recommendations, but also a relevant part of indirect trust.
\be{definition}
Suppose that we are given a trust network $\TTt = <\AAa,\EEe>$ with  
\bear
\AAa   & = &   \big([0,1] \oot{\bindd} \Bin \tto{<\partial,\varrho>} \Atr\times \Obj\big)\\
\EEe   & = &  \big([0,1] \oot{\dell} \Del \tto{<\partial,\varrho>} 
\Atr\times \Atr\big)
\eear
and moreover a trust threshold $\eta \in [0,1]$, and a composition penalty $\epsilon \in [0,1]$. The {\em endorsement completion\/} of\/ $\TTt$ is the recommendation network
\bear
\AAa^{\pthc} & = &  \big([0,1] \oot{\bindd} \Bin^\pthc \tto{<\partial,\varrho>} \Atr\times \Obj\big)\mbox{ where}\\
\Bin_{ui}^{\pthc} & = & \big\{<e,c> \in \sum_{v\in\Atr}\Del_{uv}^*\times\Bin_{vi}\ |\ \bindd(e,c)\geq \eta \}\mbox{ and}\\ 
\bindd\big(u \stackrel{e}{\rightarrow} v \stackrel{c}{\rightarrow} i \big) & 
= & \dell(e)\cdot \bindd(c)\eear
where $\Del_{uv}^*$ denotes the set of all paths in from $u$ to $v$ in\/ $\EEe$, including the empty path $\o$ if $u=v$, in which case $\dell(\o) = 1$.
\ee{definition}

\paragraph{{\bf Assumption.}} {\em In the rest of the paper, we work with recommendation networks $\AAa = \AAa^\pthc$, assumed to be endorsement complete.} 

In the next section we analyze how individual users build their own trust vectors. The repercussions of this analysis to public trust networks are discussed in section \ref{Combining}.

\section{\Indiv trust}\label{individual}
For intuition, we introduce the mathematical model of the process of trust building and updating in terms of an imaginary shopper trying out some web merchants. The model is, however, completely general, and we explain later that a recommender also builds his trust vector by an analogous process.

\subsection{{\Indiv} trust vectors and their updating}\label{updating}
The shopper records her trust in a {\em trust vector\/} $\tau \in \Rat^\Obj$. As the time $t = 0, 1,2,\ldots$ ticks, the shopper interacts with the shops, and subsequently updates $\tau$ according to her shopping experiences. This evolution makes the trust vector into a stochastic process $\tau: \NNn \to \DDD(\Rat^\Obj)$, which expresses the likely distribution of shopper's trust at time $t$ as the random variable $\tau(t) \in \DDD(\Rat^\Obj)$. The stationary distribution of the stochastic process $\tau$ is the likely distribution of trust, which we would like to analyze.

On the side of the recommenders, the shopper may also maintain a trust vector $\sigma \in \Rat^\Atr$. The idea that a trusted recommender recommends reliable merchants is expressed through the invariant $\tau_i = \sum_{u\in \Atr} \sigma_u A_{ui}$, which should be maintained as $\tau$ is updated. This makes $\sigma: \NNn\to \DDD(\Rat^\Atr)$ into another stochastic process. 


Initially, at $t=0$, the shopper may assign all merchants the same trust rating $\tau_i(0) = 1$; or she may assign each recommender the same trust rating $\sigma_u(1) = 1$, and derive $\tau_i(0) = \sum_{u\in \Atr}  A_{ui}$. 

The stochastic process $X:\NNn \to \DDD \Obj$ represents shopper's shopping history. Each random variable $X(t) \in \DDD\Obj$ selects the merchant with whom the shopper interacts at time $t$. We assume that $X(0)$ is distributed uniformly at random, whereas the probability that the next shop $X(t+1)$ will be $i\in \Obj$ is either proportional to the trust $\tau_i(t)$, or it is a fixed value $\alpha \in [0,1]$, if $i$ has had a minimal trust rating, and selecting it means replacing it by a new, untested shop. Formally, 
\bea\label{probX}
\Prob\Big(X(t+1) = i\Big) & = &\begin{cases} 
\alpha & \mbox{ if $\tau_i(t)$ was minimal (so $i$ is now new)  }\\
C(t)\tau_i(t) & \mbox{ otherwise}
\end{cases}
\eea
where $C(t) = \frac{1-\alpha}{\sum_{i\in \Obj} \tau_i(t)}$ is the normalization factor. The minimality of $\tau_i(t)$ means that for all $j\in \Obj$ holds $\tau_i(t)\leq \tau_j(t)$. The $\alpha$-case corresponds to shopper's habit to, every once in a while replace an untrusted shop, with a minimal rating, with a new, untested shop.

After the transaction with the merchant $X(t+1)$, the shopper updates her trust vector $\tau(t)$ to $\tau(t+1)$, depending on whether the merchant acted honestly or not:
\bear
\tau_{i} (t+1) & = & \begin{cases} \tau_{i} (t) & \mbox{ if $i\neq X(t+1)$}\\
0 & \mbox{ if $i = X(t+1)$ is dishonest}\\
1 & \mbox{ if $i = X(t+1)$ is honest, and new (i.e., $\tau_{i}(t)$ was minimal)}\\
1+\tau_{i} (t) & \mbox{ if $i = X(t+1)$ is honest, not new (i.e.,$\tau_{i}(t)$ not minimal)}
\end{cases}
\eear
The interpretation of the third case is that the label $i = X(t+1)$ is reassigned from some untrusted merchant, which had a minimal trust rating $\tau_i(t)$, to a new merchant, whose initial trust rating is set to $1$ if the initial transaction with was satisfactory. In the fourth case, the merchant $i =X(t+1)$ was tried out before, and has accumulated a trust rating $\tau_{X(t+1)}$, which is now increased to $\tau_{X(t+1)}(t+1) = 1+\tau_{X(t+1)} (t)$ because of a satisfactory transaction.

\subsection{{\Indiv} trust distribution}
If the trust ratings evolve according to the process just described, how will they, in the long run, partition the set $\Obj$ of merchants? How many merchants will there be with a trust rating of 1, how many with a trust rating of 2, and so on? More precisely, we want to estimate the likely number of elements in each of the sets $W_\ell(t) = \{i\in \Obj\ |\ \tau_i(t) = \ell\}$, for $\ell\in \Rat$, as the time $t$ ticks ahead. So we set up a system of equations, describing the evolution of
\bear
w_\ell(t) &=& | \{i\in \Obj\ |\ \tau_i(t) = \ell\} | 
\eear
where $|{\sf Y}|$ denotes the number of elements of the set $\sf Y$. Note that the disjoint union is $\cup_{\ell\in \Rat} W_\ell(t) =  \Obj$, and therefore $\sum_{\ell \in \Rat} w_\ell(t) = J$, where we write $J=|\Obj|$.

The initial values $w_\ell(0)$ are determined by shopper's choice of $\tau(0)$. If she sets $\tau_i(0)= 1$ for all $i\in \Obj$, then $w_1(0) = J$. 

How does $w_1$ change at the time $t$? We claim that
\bear
w_1(t+1)- w_1(t) & = &\ \ \ \ J\cdot \Prob\big(X(t+1) = i\ |\ \tau_{i}\mbox{ minimal}\big)\cdot \gamma_\bot \\
&&  -\  w_1(t) \cdot \Prob\big(X(t+1) = i\ |\ \tau_i(t)=1\big) \\
& = & J\alpha\gamma_\bot -  w_1(t)\cdot C(t)
\eear
To justify this, note that the difference between $W_1(t+1)$ and $W_1(t)$ comes about for one of the two reasons: 
\begin{itemize}
\item either $i\in \Obj$ is added to $W_1(t)$, because $\tau_i(t)$ was minimal, and $X(t+1) = i$ was selected, with the probability $\alpha$ to be replaced with a new shop from $\Obj$; and then that new shop, now called $i$, provided an honest transaction, the probability of which is $\gamma_\bot$; so $i$ is now assigned the trust rating $\tau_i(t+1)=1$;
\item or $i\in \Obj$ is deleted from $W_1(t)$, because $\tau_i(t)$ was 1, and $X(t+1) =i$ was selected from $W_1(t)$, with the probability $C(t)\cdot \tau_i(t)$; after the transaction, $i$'s trust rating was updated either to $\tau_i(t+1) = 2$ or to $\tau_i(t+1)=0$, depending on whether he acted honestly or dishonestly; but $i$ was deleted from $W_1(t)$ in any case.
\end{itemize}

However, when the ratings $\ell\gt 1$ are updated, it will not be irrelevant whether $i$ acts honestly or dishonestly. To describe dynamics of this process, we denote by $\gamma_\ell\in [0,1]$ the probability that a shop with a rating $\ell$ is honest. With the described process of trust updating, accumulating a high trust rating $\ell$ takes time. In order to get a high trust rating, a dishonest shop has to act honestly for a long time. It is therefore reasonable to assume that the probability $1-\gamma_\ell$ that an $\ell$-rated shop is dishonest decreases to 0 as $\ell$ increases; i.e. that $\lim_{\ell\rightarrow \infty} \gamma_\ell = 1$.  

Rating dynamics is now 
\bear
w_\ell(t+1)- w_\ell(t) & = &   w_{\ell -1}(t) \cdot \Prob\big(X(t+1) = i\ |\ \tau_{i}(t)=\ell -1\big)\cdot \gamma_{\ell-1}  \\
&& - w_\ell(t) \cdot \Prob\big(X(t+1) = i\ |\ \tau_i(t)=\ell\big)  \\
& = & w_{\ell -1}(t) \cdot C(t)\cdot (\ell-1) \cdot \gamma_{\ell-1}  - w_\ell (t) \cdot C(t)\cdot \ell 
\eear
The difference between $W_\ell(t+1)$ and $W_\ell(t)$ again comes from two sources: 
\begin{itemize}
\item either $i\in \Obj$ is added to $W_\ell (t)$, because $\tau_i(t)$ was $\ell -1$ and $X(t+1) = i$ was selected from $W_{\ell-1}(t)$ with the probability $C(t)\cdot (\ell-1)$; and then this $i$ turned out to be honest, with the probability $\gamma_{\ell-1}$, so that $\tau_i(t+1)$ got updated to $1+ \tau_i(t) = \ell$;
\item or $i\in \Obj$ is deleted from $W_\ell(t)$, because $\tau_i(t)$ was $\ell$, and $X(t+1) =i$ was selected from $W_\ell(t)$, with probability $C(t)\cdot \ell$; if $i$ acted honestly, his trust rating got updated to $\ell+1$; if he acted dishonestly, it got updated to 0; in any case, he got removed from $W_\ell(t)$.
\end{itemize}
Conceptually, the above derivations follow Simon's master equation method \cite{SimonH:skew}. To simplify the solution, we use a more contemporary approach of \cite{Norris,Wormald}. First of all, we do not  seek the solutions for the sizes $w_\ell(t)$ of the sets $W_\ell(t)$, but rather for the densities $v_\ell(t) = \frac{w_\ell(t)}{J}$. Since $\sum_{\ell\in \Rat} v_\ell(t) = 1$, for every $t$, the functions $v_{(-)}(t):\Rat \to [0,1]$ are probability distributions with a finite support. Together, they thus form a stochastic process $v:\NNn\to \DDD\Rat$, described by the difference equations
\bear
\Delta v_1(t) & = & \alpha\gamma_\bot  - C(t)v_1(t)\\
\Delta v_\ell(t) & = & \gamma_{\ell-1} (\ell-1) C(t) v_{\ell -1}(t)  -  \ell C(t) v_\ell (t)
\eear
As shown in the Appendix, the steady state of this process turns out to be
\[
\upsilon_1 = \frac{\alpha\gamma_\bot}{c+1}\qquad \qquad \upsilon_n  = \frac{\alpha\gamma_\bot G_{n-1}}{c} B\left(n,1+\frac{1}{c}\right)
\]
where $G_n = \prod_{\ell =1}^n \gamma_\ell$, the constant $c$ satisfies $\frac{c}{t}\approx C(t) = \frac{1-\alpha}{\sum_{i\in \Obj} \tau_i(t)}$, and  $B$ is Dirichlet's Beta function. But Stirling's formula implies that $B(x,y) \approx x^{-y}$ holds as $x\rightarrow \infty$. We have thus proven that, with a sufficiently fine trust rating scale, and with the probability of honesty $\gamma_\ell$ increasing with  the trust rating $\ell$ fast enough, the trust ratings obey the power law \cite{mitzenmacher04history,newman05zipf}.

In summary, we have proven the following:
\begin{onlythm}\label{powlow}
A trustor maintains trust ratings for a set of $J$ trustees. The ratings take their values from a sufficiently large set, so that they can strictly increase whenever justified. They are updated according to the following procedure:
\begin{itemize}
\item Initially, the tustor assigns some fixed ratings (e.g., equal) to all trustees.
\item Then the trustor repeatedly tests the trustees:
\begin{itemize}
\item with a probability $\alpha$, she tests an untested trustee, adds it to the set $J$, and deletes from it a trustee with the minimal rating;
\item otherwise, the turstor tests a previously tested trustee, with a probability proportional to its trust rating.
\end{itemize}
\item After each step, the trustor updates the trust rating $\ell$ of the tested trustee as follows
\begin{itemize}
\item with a probability $\gamma_\ell$, she increases it (because of a satisfactory outcome of the test);
\item otherwise, she sets it to zero.
\end{itemize}
\end{itemize}
If the probability $\gamma_\ell$ of a satisfactory transaction with an $\ell$-rated trustee increases fast enough enough to satisfy $\frac{1}{e^{s_\ell}}\leq \gamma_\ell \leq 1$ for some convergent series $\sum_{\ell = 1}^\infty s_\ell \lt \infty$, so that $G = \prod_{\ell = 1}^\infty \gamma_\ell\gt 0$, then in the long run, the number $w_n$ of trustees with the trust rating $n$ obeys the power law
\bear
w_n  & \approx &  \frac{\alpha\gamma_\bot GJ}{c}\  n^{-\left(1+\frac{1}{c}\right)}
\eear
where $c$ is a renormalising constant $c\approx \frac{1-\alpha}{1+\alpha\gamma_\bot}$, and $\gamma_\bot$ is the probability that an untested trustee will satisfy the test.
\end{onlythm}

\paragraph{Remarks.}  As explained in section \ref{Recommendation}, the assumption that the trust can always increase does not mean that the trust ratings have to be unbounded: they can also increase asymptotically. This assumption is only needed to assure that the process of trust building will not become irrelevant after some threshold is reached. In reality, of course, only finitely many interactions with finitely many shops can be taken into account, but there is a real sense in which the trust process can always be refined, and trust increased. 
 
The assumption that $G = \prod_{\ell = 1}^\infty \gamma_\ell\gt 0$ means that the probability $1-\gamma_\ell$, that a shop with a trust rating $\ell$ is not trustworthy, quickly decreases as $\ell$ increases. This assumption is not satisfied if many untrustworthy shops act honestly for a long time, waiting to accumulate trust, and then strike. If there are incentives for that, the heavy tail of the power component of $w_n$ is trimmed by the exponential component $G_n = \prod_{\ell = 1}^n \gamma_\ell$, and the distribution of trust is exponential. 

But this leads to a negative feedback: as they decrease the range of trust distribution, the dishonest trust hubs actually  decrease the vulnerability of the network. The more persistent attackers there are, the higher the cost of an attack.

\subsubsection{Other interpretations.} Although our model was described and motivated as shopper's trust process, it seems likely that the stochastic process governing recommender's trust vector would be of the same type. The main difference is, of course, that the recommender does not select and test the merchant himself, but builds his trust vector from the merchant ratings that he obtains as the feedback from the shoppers. However, a shopper who comes back to submit the feedback is probably the same one who previously came to obtain recommender's recommendation. And it is furthermore just as likely that the shopper has selected the merchant following that recommendation. So the selection of the merchant whose trust rating will be updated at a time $t+1$ was guided by recommender's trust vector at time $t$, just as it was the case with shopper's trust dynamics.

\subsection{Robustness and vulnerability of {\indiv} trust}\label{Robustness}
The upshot of the Theorem just proved is that there is a great variety of trust ratings: the distribution has a heavy tail. Money attracts money, and trust attracts more trust. As you extend the circle of merchants and the rating scale, you will find merchants with higher and higher trust rating. This applies to user's private trust vectors $\tau$ and $\sigma$, as well as to recommender's public trust vectors, displayed as the rows of the recommendation matrix $A = (A_{ui})_{\Atr\times\Obj}$. Moreover, although we did not describe dynamics of an endorsement network here, it seems certain that it also leads to a distribution of recommenders' influence, obeying the power law. The reason is that the endorsement dynamics is quite similar to promotion dynamics, described in \cite{PavlovicD:CSR08}, which is a version of one of the processes studied in Simon's seminal paper about the power law \cite{SimonH:skew}.

The structure and the properties of the distributions that obey the power law have been extensively analyzed \cite{mitzenmacher04history,newman05zipf,Newman:networks-book}. As mentioned in the Introduction, because of the presence of highly rated hubs, such distributions tend to be robust under random perturbations, but vulnerable to adaptive attacks on their hubs\footnote{One way to make this statement precise is to build a random graph with the given trust distribution as the degree distribution. The methods of \cite{Aiello} can serve for this purpose. The edges of the obtained graph can be interpreted as the interactions recorded in nodes' trust ratings. The trust hubs would then be the graph hubs in the usual sense: highly connected nodes. The robustness would manifest itself as a high phase transition: the graph remains connected even when many randomly selected edges are eliminated; and the fragility would mean that the graph falls apart very easily if some of the hubs are removed.}. Leaving the mathematical details aside, the security consequence is that {\em the power law distributions work for the attacker\/}: he only needs to attack a small number of nodes of high ranking, in order to gain control over a large part of the system. This phenomenon has been previously demonstrated on toy models of trust networks, involving the bottleneck nodes \cite{LevienR}. Although the recommender networks, currently deployed on the Web, still do not form a large network, the same phenomenon --- that the main trust hubs become increasingly unreliable --- has also been observed in practice: e.g., \cite{EdelmanB:adverse} describes some extreme examples. 

\section{Conclusions}\label{Conclusions}
The obvious security lessons, arising from our analyses, and supported by the empiric observations are thus: 
\begin{itemize}
\item Trust decisions should not be derived from public trust recommendations alone. They should be based on private trust vectors, that the user should maintain herself.
\item Public trust recommendations should be used to supplement and refine private trust.
\end{itemize}

\section{Towards applications:\\
Combining private trust and public recommendations}\label{Combining}
Hoping that the gentle reader will not be too disturbed by the fact that the paper continues beyond its conclusions\footnote{A reviewer of a version of this paper where the above conclusions were not separated in their own section, objected that the paper ended abruptly, without any conclusions.}, in this final section we sketch some ways to implement these conclusions. We propose for further exploration two methods for a user of a trust network to combine her private trust vectors with some public recommendations, in order to obtain more informative trust guidance. Although we attempt to provide intuitive explanations, understanding the technical details of these condensed ideas may require some familiarity with LSI and with the vector model.

\subsection{Trust communities}
It is often emphasized that trust is relative to a community, or more generally to a module \cite{PavlovicD:CSR08} within a network: e.g., a criminal may be trusted within the community of criminals, but not within a community of security researchers, and vice versa. The members of the same community can be recognized by similar trust vectors, or recommendations.

In this section, we briefly summarize how a recommendation matrix can be used to recognize communities in the space of recommenders on one hand, and in the space of merchants on the other. The merchants which deserve to be trusted for the same type of services are likely to be highly recommended by the same recommenders. This groups them into communities. The user  can refine his trust by computing how much he trusts each community, and how is his trust  distributed within each of them. While the public trust recommendations may be unreliable, and better not followed directly, they provide reliable and valuable information about the trust communities. By relativizing the private trust over the trust communities, the user can obtain significantly more precise guidance, distinguishing between the various forms of trust in the various communities, even in the model where the entrusted properties are kept implicit. 

By suitably renormalizing the data, the similarity between the trust vectors $\varphi$ and $\psi\in \Rat^{\Obj}$ can be viewed as the angle between the induced recommender vectors
\bear
\Sim(\varphi,\psi) & = & <A\varphi\ |\ A\psi>
\eear
where $<x|y> = \sum_{v\in\Atr} x_v\cdot y_v$ is the inner product in the space $\Rat^{\Atr}$. The angle is often used as the similarity measure in information retrieval and data mining \cite{Raghavan:IR-book}. It should be noted that it leads to subtle statistical problems, if applied to diverse samples \cite{PavlovicD:QI08}. 
The trust communities, as the subspaces of similar vectors within $\Rat^{\Obj}$, can be detected by spectral methods, using the data mining technique of Latent Semantic Indexing (LSI) \cite{LSI,kleinberg99authoritative,PavlovicD:QI08}. The idea is to look for the vectors $\xi$ where $\Sim(\xi,\xi)$ attains the extremal values. Since the transpose $A^T$ satisfies $<A\varphi\ |\ A\psi> = <\varphi\ |\ A^T A\psi>$, the similarity can be also be expressed as $\Sim(\varphi,\psi) =  <\varphi\ |\ A^T A\psi>$. The extremal values of $\Sim(\xi,\xi) =  <\xi\ |\ A^T A\xi>$ can thus be found as the eigenvalues $\{\lambda_1\gt \lambda_2\gt\cdots\gt \lambda_m\}$  of $A^T A$. The communities are the corresponding eigenspaces, described by the projectors $\{P_1,\ldots, P_m\}$. 

There are at least two ways to refine private trust $\tau$ using the trust communities $\{P_1,\ldots,P_m\}$. 

\subsubsection{Community specific private trust.}
Instead of using his trust vector $\tau\in \Rat^{\Obj}$ to select the trusted objects, the user can compute the {\em community specific\/} trust vectors
\bear
\tau^k & = & P_k \tau
\eear
obtained by projecting $\tau$ into each of the eigenspaces $P_k$, $k=1,\ldots,m$, i.e. by relativizing it to the dominant merchant communities. In this way, even if the trust relations $A\truss{}{}{} B$ are not explicitly annotated by the entrusted properties $\Phi$, the user can refine his trust decisions by recognizing the "latent" entrusted properties,  uncovered as the dominant trust communities $\{P_1,\ldots,P_m\}$.

\subsubsection{Personalized recommendation matrix.}
Intuitively, the spectrum $\{\lambda_1\gt \lambda_2\gt\cdots\gt \lambda_m\}$ expresses a notion of cohesion, i.e. the strength of the mutual trust within each of the communities 
$\{P_1,P_2,\ldots,P_m\}$. On the other hand, the degree to which a user with a trust vector $\tau$ trusts a community $P_k$ can be measured by the similarity 
$\Sim(\tau,\tau^k) = <\tau\ |\ P_k\tau>$.

The Singular Value Decomposition (SVD) theorem tells that the spectral decomposition $A^T A = \sum_{k=1}^m \lambda_k P_k$ induces $A = \sum_{k=1}^m \sqrt{\lambda_k} \Pi_k$, for the suitable operators $\Pi_k$. The personalized recommendation matrix, remixed according to the community trust $\theta$ induced by user's trust vector $\tau$ is then $A_\tau = \sum_{k=1}^m \sqrt{
<\tau |P_k\tau>} \Pi_k$. Using this private matrix is equivalent to using the community specific trust vectors, within each of the trust communities; but it also allows evaluating trust for  combinations of communities.

\bibliographystyle{plain}
\bibliography{ref-trust,ref-trust-1,PavlovicD}

\begin{thebibliography}{10}

\bibitem{Aiello}
William Aiello, Fan Chung, and Linyuan Lu.
\newblock A random graph model for massive graphs.
\newblock In {\em STOC '00: Proceedings of the thirty-second annual ACM
  symposium on Theory of computing}, pages 171--180, New York, NY, USA, 2000.
  ACM.

\bibitem{Benantar:AC}
Messaoud Benantar.
\newblock {\em Access Control Systems: Security, Identity Management and Trust
  Models}.
\newblock Springer Verlag, 2006.

\bibitem{BBK}
Thomas Beth, Malte Borcherding, and Birgit Klein.
\newblock Valuation of trust in open networks.
\newblock In {\em ESORICS '94: Proceedings of the Third European Symposium on
  Research in Computer Security}, pages 3--18, London, UK, 1994.
  Springer-Verlag.

\bibitem{BlazeM:Decentralized}
Matt Blaze, Joan Feibenbaum, and Jack Lacy.
\newblock Decentralized trust management.
\newblock In {\em Proceedings of Symposium on Security and Privacy}, page 164,
  1996.

\bibitem{SassoneV:Trust}
Marco Carbone, Mogens Nielsen, and Vladimiro Sassone.
\newblock A formal model for trust in dynamic networks.
\newblock In A.~Cerone and P.~Lindsay, editors, {\em Proceedings of the First
  International Conference on Software Engineering and Formal Methods}, 2003.

\bibitem{Norris}
R.W.R. Darling and James~R. Norris.
\newblock Differential equation approximations for {Markov} chains.
\newblock {\em Probability Surveys}, 5:37--79, 2008.

\bibitem{LSI}
Scott~C. Deerwester, Susan~T. Dumais, Thomas~K. Landauer, George~W. Furnas, and
  Richard~A. Harshman.
\newblock Indexing by latent semantic analysis.
\newblock {\em Journal of the American Society of Information Science},
  41(6):391--407, 1990.

\bibitem{EdelmanB:adverse}
Benjamin Edelman.
\newblock Adverse selection in online trust certifications.
\newblock working paper, {\tt
  http://www.benedelman.org/publications/advsel-trust-draft.pdf}.

\bibitem{Guha-Tomkins}
R.~Guha, Ravi Kumar, Prabhakar Raghavan, and Andrew Tomkins.
\newblock Propagation of trust and distrust.
\newblock In {\em WWW '04: Proceedings of the 13th international conference on
  World Wide Web}, pages 403--412, New York, NY, USA, 2004. ACM.

\bibitem{GuttmanJ:Trust}
Joshua~D. Guttman, F.~Javier Thayer, Jay~A. Carlson, Jonathan~C. Herzog,
  John~D. Ramsdell, and Brian~T. Sniffen.
\newblock Trust management in strand spaces: A rely-guarantee method.
\newblock In David~A. Schmidt, editor, {\em ESOP}, volume 2986 of {\em Lecture
  Notes in Computer Science}, pages 325--339. Springer, 2004.

\bibitem{JosangA:Subjlog}
Audun J{\o}sang.
\newblock A subjective metric of authentication.
\newblock In {\em ESORICS '98: Proceedings of the 5th European Symposium on
  Research in Computer Security}, pages 329--344, London, UK, 1998.
  Springer-Verlag.

\bibitem{JosangA:PGP}
Audun J{\o}sang.
\newblock An algebra for assessing trust in certification chains.
\newblock In {\em NDSS}. The Internet Society, 1999.

\bibitem{JosangA:Trustnet06}
Audun J{\o}sang, Elizabeth Gray, and Michael Kinateder.
\newblock Simplification and analysis of transitive trust networks.
\newblock {\em Web Intelli. and Agent Sys.}, 4(2):139--161, 2006.

\bibitem{Garcia-Molina:eigentrust}
Sepandar~D. Kamvar, Mario~T. Schlosser, and Hector Garcia-Molina.
\newblock The {Eigentrust} algorithm for reputation management in {P2P}
  networks.
\newblock In {\em WWW '03: Proceedings of the 12th international conference on
  World Wide Web}, pages 640--651, New York, NY, USA, 2003. ACM Press.

\bibitem{Karabulut}
Y{\"u}cel Karabulut, Florian Kerschbaum, Fabio Massacci, Philip Robinson, and
  Artsiom Yautsiukhin.
\newblock Security and trust in it business outsourcing: a manifesto.
\newblock {\em Electr. Notes Theor. Comput. Sci.}, 179:47--58, 2007.

\bibitem{kleinberg99authoritative}
Jon~M. Kleinberg.
\newblock Authoritative sources in a hyperlinked environment.
\newblock {\em Journal of the ACM}, 46(5):604--632, 1999.

\bibitem{LampsonB:AC}
Butler Lampson, Mart\'{\i}n Abadi, Michael Burrows, and Edward Wobber.
\newblock Authentication in distributed systems: theory and practice.
\newblock {\em SIGOPS Oper. Syst. Rev.}, 25(5):165--182, 1991.

\bibitem{Langville06google-book}
Amy~N. Langville and Carl~D. Meyer.
\newblock {\em Google's {PageRank} and Beyond: The Science of Search Engine
  Rankings}.
\newblock Princeton University Press, Princeton, NJ, USA, 2006.

\bibitem{LevienR}
Raph Levien and Alexander Aiken.
\newblock Attack-resistant trust metrics for public key certification.
\newblock In {\em SSYM'98: Proceedings of the 7th conference on USENIX Security
  Symposium, 1998}, pages 18--18, Berkeley, CA, USA, 1998. USENIX Association.

\bibitem{Ninghui:Trust}
Ninghui Li, John~C. Mitchell, and William~H. Winsborough.
\newblock Design of a role-based trust-management framework.
\newblock In {\em SP '02: Proceedings of the 2002 IEEE Symposium on Security
  and Privacy}, page 114, Washington, DC, USA, 2002. IEEE Computer Society.

\bibitem{Ninghui:JACM}
Ninghui Li, John~C. Mitchell, and William~H. Winsborough.
\newblock Beyond proof-of-compliance: security analysis in trust management.
\newblock {\em J. ACM}, 52(3):474--514, 2005.

\bibitem{Raghavan:IR-book}
Christopher~D. Manning, Prabhakar Raghavan, and Hinrich SchŸtze.
\newblock {\em Introduction to Information Retrieval}.
\newblock Cambridge University Press, Cambridge, UK, 2008.

\bibitem{Garcia-Molina:Taxonomy}
Sergio Marti and Hector Garcia-Molina.
\newblock Taxonomy of trust: categorizing {P2P} reputation systems.
\newblock {\em Comput. Netw.}, 50(4):472--484, 2006.

\bibitem{MaurerU:Trust}
Ueli Maurer.
\newblock Modelling a public-key infrastructure.
\newblock In {\em {ESORICS}: European Symposium on Research in Computer
  Security}. LNCS, Springer-Verlag, 1996.

\bibitem{mitzenmacher04history}
Michael Mitzenmacher.
\newblock A brief history of generative models for power law and lognormal
  distribution.
\newblock {\em Internet Math.}, 1:226--251, 2004.

\bibitem{newman05zipf}
Mark Newman.
\newblock Power laws, {Pareto} distributions and {Zipf's} law.
\newblock {\em Contemporary Physics}, 46:323, 2005.

\bibitem{Newman:networks-book}
Mark Newman, Albert-Laszlo Barabasi, and Duncan~J. Watts, editors.
\newblock {\em The Structure and Dynamics of Networks}.
\newblock Princeton Studies in Complexity. Princeton University Press,
  Princeton, NJ, USA, 2006.

\bibitem{PavlovicD:CSR08}
Dusko Pavlovic.
\newblock Network as a computer: ranking paths to find flows.
\newblock In Alexander Razborov and Anatol Slissenko, editors, {\em Proceedings
  of Third International Computer Science Symposium in Russia}, volume 5010 of
  {\em Lecture Notes in Computer Science}, pages 384--397. Springer Verlag,
  2008.

\bibitem{PavlovicD:QI08}
Dusko Pavlovic.
\newblock On quantum statistics in data analysis.
\newblock In Peter Bruza, editor, {\em Quantum Interaction 2008}. AAAI, 2008.

\bibitem{ReiterM:Metric}
Michael~K. Reiter and Stuart~G. Stubblebine.
\newblock Authentication metric analysis and design.
\newblock {\em ACM Trans. Inf. Syst. Secur.}, 2(2):138--158, 1999.

\bibitem{SimonH:skew}
Herbert~A. Simon.
\newblock On a class of skew distribution functions.
\newblock {\em Biometrika}, 42:425--440, 1955.

\bibitem{Watts-Strogatz}
D.~J. Watts and S.~H. Strogatz.
\newblock Collective dynamics of 'small-world' networks.
\newblock {\em Nature}, 393(6684):440--442, June 1998.

\bibitem{Wormald}
Nicholas~C. Wormald.
\newblock Differential equations for random processes and random graphs.
\newblock {\em The Annals of Applied Probability}, 5(4):1217--1235, 1995.

\end{thebibliography}

\appendix
\section*{Appendix: The steady state of the trust process}
The trust process $v:\NNn\to \DDD\Rat$ is described by the difference equations
\bear
\Delta v_1(t) & = & \alpha\gamma_\bot  - C(t)v_1(t)\\
\Delta v_\ell(t) & = & \gamma_{\ell-1} (\ell-1) C(t) v_{\ell -1}(t)  - C(t) \ell v_\ell (t)
\eear
Recall, first of all, from section \ref{updating} that $C(t) = \frac{1-\alpha}{S(t)}$, where $S(t) = \sum_{i\in \Obj} \tau_i(t)$. The dynamics of $\tau$, described at the end of section \ref{updating}, implies that 
\[
S(t+1) = \sum_{i\neq X(t+1)} \tau_i(t) + \gamma_{X(t+1)} \left(1+ \tau_{X(t+1)}(t)\right) +\alpha \gamma_\bot
\]
where $\gamma_\bot$ is the probability that a shopper is satisfied after an interaction with a new shop. It follows that 
\[
\Delta S(t) = \gamma_{X(t+1)} - (1-\gamma_{X(t+1)})\tau_{X(t+1)}(t) + \alpha \gamma_\bot \approx 1 +\alpha \gamma_\bot
\]
is approximately constant and thus $S(t) \approx (1 +\alpha \gamma_\bot)t$. Hence $C(t) \approx \frac{c}{t}$, where $c = \frac{1-\alpha}{1 +\alpha \gamma_\bot}$. 

With this simplification, and with the martingale assumption of \cite{Wormald} satisfied, the solutions of the above system of difference equations can be approximated by the solutions of the corresponding differential system
\bear
\frac{dv_1}{dt} & = & \alpha\gamma_\bot  - \frac{c}{t}v_1\\
\frac{dv_\ell}{dt} & = & \frac{\gamma_{\ell-1} c(\ell-1) v_{\ell -1}  - c \ell v_\ell}{t}
\eear
where the discrete time variable $t$ has been made continuous. The steady state of the stochastic process $v:\RRr\to \DDD\Rat$ can now be found in the form $v_\ell(t) = t\cdot \upsilon_\ell$, by expanding the recurrence
\bear
\upsilon_1 & = & \alpha\gamma_\bot  - c\upsilon_1\\
\upsilon_\ell & = & \gamma_{\ell-1} c (\ell-1) \upsilon_{\ell -1}  - c \ell \upsilon_\ell
\eear
into
\bear
\upsilon_1 & = & \frac{\alpha\gamma_\bot}{c+1}\\
\upsilon_\ell & = & \frac{(\ell-1)\gamma_{\ell-1} c}{\ell c +1}\  \upsilon_{\ell-1}
\eear
which further gives
\bear
\upsilon_2 & = & \frac{\alpha\gamma_\bot}{c+1}\cdot \frac{\gamma_1 c}{2c +1}\\
\upsilon_3 & = & \frac{\alpha\gamma_\bot}{c+1}\cdot \frac{\gamma_1 c}{2c +1}\cdot \frac{2\gamma_2 c}{3c +1}\\
& \ldots & \\
\upsilon_n & = & \alpha\gamma_\bot \left(\prod_{\ell = 1}^{n-1} \gamma_\ell\right) c^{n-1}\cdot \frac{(n-1)!}{\prod_{k = 1}^n (kc+1)}\\
& = & \frac{\alpha\gamma_\bot G_{n-1}}{c} \cdot \frac{(n-1)!}{\prod_{k = 1}^n \left(k+\frac{1}{c}\right)} \\
& = & \frac{\alpha\gamma_\bot G_{n-1}}{c} \cdot \frac{\Gamma(n)\Gamma\left(1+\frac{1}{c}\right)}{\Gamma\left(n+1+ \frac{1}{c} \right)} \\
& = & \frac{\alpha\gamma_\bot G_{n-1}}{c} \cdot B\left(n,1+\frac{1}{c}\right)
\eear

\end{document}